\documentclass[twocolumn,twoside,slac_two]{revtex4}
\usepackage{graphicx}
\usepackage{fancyhdr}
\begin{document}

\title{X--Ray view of Misaligned AGNs}

%

\author{E. Torresi}
\affiliation{INAF/IASF Bologna, via Gobetti 101, 40129 Bologna, Italy}

\begin{abstract}
The {\it Fermi}--LAT satellite has recently discovered a small group of radio galaxies and steep spectrum radio sources: the misaligned AGNs (MAGNs) sample.
We present the X--ray analysis of all the sources of this sample (7 FRIs and 3 FRIIs) with a firm GeV association. 
This study supports the idea that FRIIs host more efficient accretion mechanisms ($\dot{m} >$0.1) than FRIs ($\dot{m} <$0.003). 
Furthermore, in objects with high accretion rates the Broad Line Regions appear to be very active zones where, in addition to optical lines, the fluorescence iron K$\alpha$ feature at 6.4~keV is also produced. It seems that the FRII jets propagate in an environment very rich in photons, explaining, at least at zeroth order, why the External Compton is the preferred mechanism to produce $\gamma$--rays. In FRIs, where also the iron line is difficult to be detected,  the paucity of photons in the circumnuclear ambient seems to favor the Synchrotron Self Compton process. 
\end{abstract}

\maketitle

\thispagestyle{fancy}


\section{INTRODUCTION}
With the term misaligned AGNs (MAGNs) we define radio--loud (RL) sources whose jet is pointed away from the observer, i.e., without strong Doppler boosting, with steep radio spectra ($\alpha_{\rm r} >$0.5) and resolved and possibly symmetrical radio structures. Given the large inclination angle of the jet, MAGNs were not considered favorite GeV targets, but the {\it Fermi} Large Area Telescope (LAT; \cite{atwood}) detection of 11 MAGNs in 15 months of survey \cite{magn} has changed this view and confirmed them as a new interesting class of $\gamma$--ray emitters.
MAGNs are mainly FRI and FRII sources (and in minor part Steep Spectrum Radio Sources, SSRS, and Compact Steep Sources, CSS). 
FRIs and FRIIs have been historically classified by Fanaroff \& Riley \cite{fr} on the base of their extended radio morphology that changes over, or under, a critical radio power at 178 MHz (P$_{\rm 178~MHz} \sim$10$^{25}$~W~Hz$^{-1}$~sr$^{-1}$). 
At zeroth order, the Unified Models of AGNs \cite{urry} identify FRIs and FRIIs with the parent population of BL Lacs and Flat Spectrum Radio Quasars (FSRQs), i.e., their misaligned counterpart. 
In the optical band, radio sources are divided into Broad Line Radio Galaxies (BLRGs) and Narrow Line Radio Galaxies (NLRGs) depending on their bright (weak) optical continuum and broad (narrow) emission lines. BLRGs at high redshift with steep radio spectra ($\alpha_{\rm r} >$0.5) are usually defined as Steep Spectrum Radio Sources (SSRS). Compact Steep Sources (CSS) have also steep radio spectra, but are very small, with sizes smaller than their host galaxies.
NLRGs are further divided into High Excitation Galaxies (HEG) and Low Excitation Galaxies (LEG) \cite{jackson} \footnote{A source is classified as LEG if EW$_{\rm [OIII]}<$10~\AA\  and/or O[II]/O[III]$>$1.}. BLRGs and NLRG/HEGs have exclusively FRII morphology, while NLRG/LEGs can assume both FRI and FRII morphologies.\\
Despite the large amount of information on the properties of these objects, the cause of the FRI/FRII dichotomy is still unclear. Several scenarios have been proposed. For example, the dichotomy could be due to the different ambient medium in which the jet interacts \cite{bicknell}. In this framework FRI jets start highly relativistic and decelerate between the sub--pc and kpc scales. According to \cite{reynolds}  the dichotomy could be ascribed to differences in the jet content. Finally, \cite{ghisellini} proposed that the accretion process itself might play a key role in driving the dichotomic behavior of these sources.\\
There are several studies in the optical and infrared bands pointing out a possible difference in FRIs and FRIIs accretion modes. \cite{chiaberge} demonstrated that the optical flux of FRI radio galaxies is strongly correlated with that of the radio core over four decades, suggesting a non--thermal synchrotron origin from the base of the jet for both emissions (their Figure 2). On the contrary, the excess in the optical luminosity of FRIIs can be related to the accretion disk.  {\it HST} images of FRIs have shown that there is no nuclear absorption,  meaning that the observed weakness of the optical lines can not be ascribed to obscuration, for example, by a dusty torus.\\
IR observations support this scenario. Recently, \cite{baldi} found that the near infrared (NIR) luminosity is also correlated with the radio one (their Figures 5 and 6). FRIIs, instead, show a large NIR excess probably of thermal origin and related to hot circumnuclear dust (the torus) heated by an unbeamed and strong nuclear continuum. 
The accretion rate bimodality leading the dichotomy was also suggested by \cite{marchesini}. In particular, they found that FRIs are characterized by low accretion rates, while high accretion rates characterize FRIIs and quasars nuclei.\\
Here we present the X--ray analysis of ten MAGNs detected in 15 months of LAT survey \cite{magn}. 
X--rays are a powerful tool to address the FRI/FRII dichotomy, since they offer the unique possibility of directly observing the central engine and studying its impact on the circumnuclear environment.

\section{THE SAMPLE}
We present the X--ray nuclear analysis of all MAGNs (7 FRIs and 3 FRIIs) with a firm LAT detection. Only PKS~0943-76, although present in the original sample, is not considered here since its $\gamma$--ray association is not secure.
The studied MAGNs  belong to the Cambridge (3CR \cite{bennett}--\cite{spinrad} and 3CRR \cite{laing}) and Molonglo (MS4) \cite{ms4a}--\cite{ms4b} radio catalogs. 
Table~\ref{tab1} lists the sources of the sample together with their redshifts, their  radio and optical classifications and the X--ray satellites ({\it Chandra}, {\it XMM--Newton}, {\it Swift}, {\it Suzaku}) used to perform the analysis.

\begin{table}[!]
\begin{center}
\caption{The MAGN sample.}
\begin{tabular}{l|c|cc|c}
\hline \textbf{Objects }   &\textbf{z}  &\textbf{Class}   &    &\textbf{Sat.}$^{a}$ \\
 &   &Radio  &Optical  &     \\  
\\ 
\hline 3C~78/NGC~1218  & 0.029 & FRI & G &C \\
 
\hline 3C~84/NGC~1275  & 0.018 & FRI & G  &X \\

\hline M87/3C~274  & 0.004 & FRI &G  &C \\

\hline NGC~6251  & 0.0247 & FRI &G &X, Sw\\

\hline PKS~0625-354  & 0.055 & FRI &G    &X\\
 
\hline Cen A & 0.0009 & FRI &G &X \\

\hline 3C~120   & 0.033 & FRI &BLRG &X \\

\hline 3C~111  & 0.0491 & FRII &BLRG  &Sz, X \\

\hline 3C~207 & 0.681 & FRII/SSRQ &- &X \\

\hline 3C~380  & 0.692 & FRII/CSS & - &C\\
\hline
\multicolumn{5}{l}{(a) C={\it Chandra}; X={\it XMM--Newton}; Sw={\it Swift}; Sz={\it Suzaku}}\\
\end{tabular}
\label{tab1}
\end{center}
\end{table}

\section{X--RAY ANALYSIS}
We reduced the archival data of the sources following standard procedures. \\
In this section we show, as an example, the analysis of three sources: a typical FRI (NGC~6251), a typical FRII (3C~111) and a peculiar radio galaxy (3C~120) with FRI morphology but characterized by an efficient accretion disk. \\

\noindent
{\bf A typical FRI: NGC~6251}\\

\begin{figure}
\includegraphics[angle=270,width=50mm]{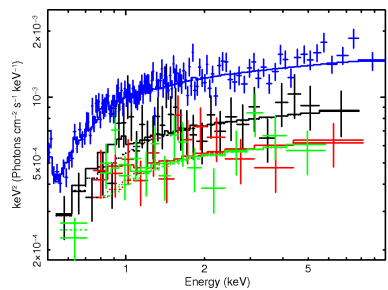}
\caption{Unfolded spectral model of the combined {\it XMM--Newton} and {\it Swift} datasets of NGC~6251. Blue points : {\it XMM--Newton} observation of 2002 March; black: {\it Swift} observation of 2007 April; red: {\it Swift} observation of 2009 May; green: {\it Swift} observation of 2009 June. For more details see \cite{6251}.}
\label{6251}
\end{figure}

NGC~6251 is a luminous FRI, the fifth $\gamma$--ray brightest object among the MAGNs (see Table~\ref{tab2}), located at z=0.0247. 
This radio galaxy, already associated with the EGRET source 3EG~J1621+8203 \cite{mukerjee},  was detected by the {\it Fermi}--LAT in the first year of survey \cite{magn}, \cite{1FGL},  \cite{1LAC}.
All available archival data, from {\it XMM--Newton} (2002 March) and {\it Swift} (2007 April -- 2009 June) satellites were re--analyzed.
\footnote{The  {\it Chandra} observation of 2003 November  was affected by pile--up therefore is not taken into account.}\\
The  {\it XMM--Newton} best--fit  spectrum (0.3--10 keV) is typical of FRI radio sources \cite{balmaverde}. It is modeled by a power law ($\Gamma$=1.89$\pm$0.04) absorbed by  Galactic column density (N$_{\rm H, Gal}$=5.4$\times$10$^{20}$~cm$^{-2}$, \cite{lab}) and intrinsic column density (a few $\times$ 10$^{20}$~cm$^{-2}$), plus a thermal component (kT$\sim$0.6~keV).  There is no significant evidence for a Fe K$\alpha$ emission line, in agreement with \cite{evans}.  \\
The {\it Swift}/X--Ray Telescope (XRT) observed NGC~6251 three times from 2007 to 2009. 
All spectral fits were performed in the 0.5--10 keV band. The best--fit model is an absorbed power law with column density slightly in excess of the Galactic value, in agreement with {\it XMM--Newton} data that require additional intrinsic absorption. The spectral slopes are steeper ($\Gamma \sim$2) than in {\it XMM--Newton} (but consistent within the errors). This is probably due to a low signal--to--noise ratio, that can not allow the detection of the soft thermal component. During the {\it XMM--Newton} pointing the source has a flux 40$\%$ higher than in the {\it Swift} observation. A further change in flux of about 15$\%$ was observed by {\it Swift} from 2007 to 2009 (see Figure~\ref{6251}). These $\approx $2 years represent the shortest period during which X--ray flux variability has been detected. This offers observational constraints on the nuclear origin of the high energy photons, probably produced in a region not larger than $\sim$1~pc. For more details about the X--ray analysis and the SED modeling refer to \cite{6251}.\\

\noindent
{\bf A typical FRII: 3C~111}\\

\begin{figure}
\includegraphics[angle=0,width=50mm]{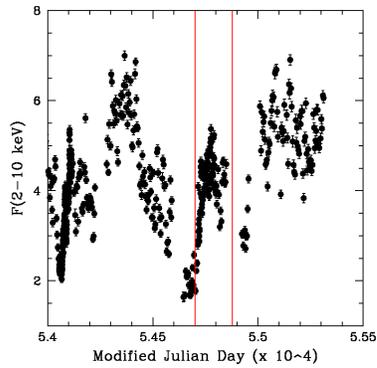}
\caption{ {\it RXTE} X--ray light curve of 3C~111 collected from 2006 September to 2010 April (from \cite{chatterjee}). The two red lines correspond to the {\it Suzaku} (2008 August) and {\it XMM--Newton} (2009 February) observations. During the first pointing the source was in an X--ray minimum, while in the second pointing, three months later, the flux was increased by a factor $\approx$2.5.}
\label{lc_x}
\end{figure}

\begin{figure}
\includegraphics[angle=270,width=45mm]{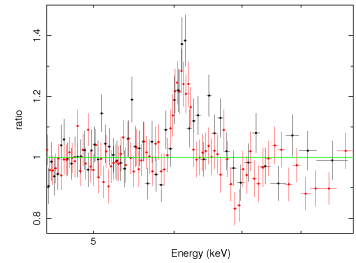}
\caption{Data--to--model ratio for the {\it Suzaku} XIS spectra after fitting the data with an absorbed power law. {\it Black} points: XIS1 (BI); {\it red} points: XIS0--3 (FI). The emission and absorption features in the iron K complex region are evident. The energy axis  refers to the observed frame.}
\label{suzaku}
\end{figure}

3C~111 (z=0.0491) is a nearby FRII radio galaxy located at low Galactic latitude. Together with Centaurus A and NGC~6251, they are the only radio galaxies considered as EGRET candidate sources \cite{sguera}--\cite{hartman}. The {\it Fermi}--LAT instrument detected a $\gamma$--ray flare from this source in the period 2008 October--November \cite{magn}--\cite{paola111}, immediately after an X--ray flux minimum (see Figure~\ref{lc_x}).\\
Here we present the analysis of two X--ray observations performed by {\it Suzaku} (t$_{\rm exp}\sim$122 ks) and  {\it XMM--Newton} (t$_{\rm exp}\sim$124 ks) on 2008 August 22  and 2009 February 15, respectively.\\
We analyzed the data of both the front--illuminated (FI) and back--illuminated (BI) X--ray Imaging Spectrometers (XIS) \cite{xis} onboard {\it Suzaku}.
The FI and BI spectra were grouped to have at least a minimum of 200 and 100 counts per bin, respectively.\\
Events within the 0.6--10 keV band were considered, and the region around the Si K edge (1.7--1.9 keV) discarded because of uncertainties in the calibration \cite{ballo}.\\
The X--ray continuum was fitted with a power law ($\Gamma \sim$1.6) absorbed by Galactic column density. We left this parameter free to vary in order to take into account the contribution of a dense molecular cloud along the source line--of--sight. 
An inspection of the residuals around the iron K complex region (5--7.5 keV) reveals the presence of features in emission and absorption (Figure~\ref{suzaku}). 
The emission line was fitted with a narrow ($\sigma$=0.07$\pm$0.03~keV) Gaussian profile of EW=75$\pm$13~eV centered at E=6.4$\pm$0.02~keV.\\
Residuals of both FI and BI data show a trough at $\sim$7.2 keV  (rest frame). The presence of such feature has been already claimed by previous authors \cite{ballo}, \cite{tombesi}. They interpreted the line as the resonant Fe {\small XXVI} Ly$\alpha$ transition at 6.97~keV, outflowing at high velocity (v$_{\rm out} \sim$10$^{4}$~km~s$^{-1}$) from the accretion disk.

\begin{figure}
\includegraphics[angle=0,width=50mm]{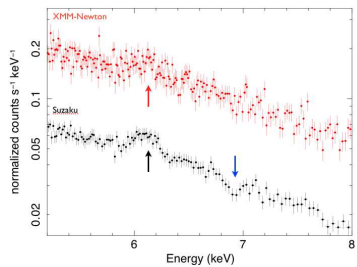}
\caption{{\it XMM--Newton} (red) and {\it Suzaku} (black) hard X--ray spectra. The difference in fluxes is evident as far as the presence of an absorption line in {\it Suzaku} but not in {\it XMM--Newton} data. The red and black arrows mark the iron K$\alpha$ line, while the blue arrow the absorption line.}
\label{suz_xmm}
\end{figure}

During the {\it XMM--Newton} observation the flux of 3C~111 was higher by a factor $\approx$2.5 (see Figures~\ref{lc_x}--\ref{suz_xmm}).\\
The spectrum is again well fitted by a power law ($\Gamma\sim$1.6) absorbed by Galactic N$_{\rm H}$ similar to that measured in the previous observation. Residuals to the power law model reveal the presence of the iron K$\alpha$ line at E=6.4$\pm$0.04~keV, characterized by $\sigma$=0.1$\pm$0.06~keV and EW=67$^{+19}_{-23}$~eV \footnote{In both {\it Suzaku} and {\it XMM--Newton} spectra the addition of a narrow Fe K$\beta$ component around 7~keV further improves the fit.}. 
There is a hint that the iron line is more prominent in the {\it XMM--Newton} observation than in the {\it Suzaku} one. The line seems to follow the continuum change but, given the large uncertainties on the line flux (and EW), no firm conclusion can be drawn.
However, a study of the iron line, based on six  years of monitoring with the {\it RXTE} satellite \cite{chatterjee}, indicates a strong correlation between the continuum and the iron line flux.\\
It is worth noting the lack in the {\it XMM--Newton} spectrum of the absorption feature around 7~keV (Figure~\ref{suz_xmm}). If these high velocity outflows really exist they should be transient phenomena. As argued by \cite{paola111} the presence of these winds in the X--ray spectrum could be related to a change in the disk--jet system.  

Using the better constrained intrinsic width of the iron line detected by {\it Suzaku} ($\sigma$=0.07$\pm$0.03~keV), we can derive the velocity width of the K$\alpha$ feature assuming Keplerian rotation around the black hole.
The obtained velocity width (FWHM) is $\sim$8000$^{+2000}_{-4000}$~km~s$^{-1}$, that is in the typical velocity range of Broad Line Regions (BLR) clouds. \\

\noindent
{\bf3C~120: a FRI with a powerful accretion disk}\\

\begin{figure}
\includegraphics[angle=270,width=40mm]{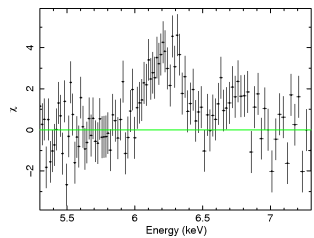}
\caption{3C~120 residuals after fitting the hard X--ray spectrum with a power law. Two prominent emission lines are visible. The best--fit values of the two Gaussian lines are: E$_{1}$=6.4$^{+0.03}_{-0.04}$~keV, $\sigma_{1}$ =0.12$^{+0.08}_{-0.04}$~keV, EW$_{1}$=62$^{+22}_{-18}$~eV; E$_{2}$=6.92$\pm$0.04~keV, $\sigma_{2}<$0.1 keV, EW$_{2}$=26$^{+21}_{-8}$~eV.}
\label{spec120}
\end{figure}

3C~120 (z=0.033; \cite{burbidge}) is the softest $\gamma$--ray source of the MAGN sample ($\Gamma_{\gamma} \sim$2.7). Its $\gamma$--ray emission was detected by {\it Fermi}--LAT in the 0.1--100~GeV energy band with a significance of $\approx$5.6 $\sigma$ \cite{magn}.\\
It is classified as FRI with a prominent radio jet lying at an angle of $\sim$20$^{\circ}$ to our line--of--sight \cite{jorstad} and showing strong flux and structure variability \cite{gomez}, \cite{walker}.  \\
Its optical spectrum is typical of Seyfert 1 galaxies with strong and broad emission lines, quite unusual for FRI radio sources. At UV wavelengths it shows a strong blue bump and strong emission lines signatures of a standard optically--thick geometrically--thin accretion disk \cite{maraschi}. 
After $\approx$3 years of X--ray and radio monitoring of 3C~120, \cite{marscher}  claimed the existence of a direct link between disk and jet, similar to that observed in black hole X--ray binaries. They found that X--ray dips are followed by the emission of bright superluminal radio knots. \cite{chatterjee2} support this claim using a longer monitoring. In addition, they found an anti--correlation between X--ray and 37 GHz fluxes, with X--rays leading the radio by $\sim$120 days. A strong correlation between X--ray and optical variations is also observed implying that optical and X--ray photons are produced in the same emitting region, probably the accretion disk--corona system.\\ 
In order to study the nuclear X--ray behavior of the source we re--analyzed an {\it XMM--Newton} archival observation of 130 ks (for further details see also \cite{ballantyne}, \cite{ogle}).
Only pn and MOS2 data (taken in Small Window mode) were considered.
Spectra were binned to have at least 20 counts in each spectral channel in order to apply the $\chi^{2}$ statistic. \\
The spectrum of 3C~120 is complex. In particular, modeling the soft X--ray continuum (0.4--3 keV) is not trivial. For this reason here we concentrate on the hard X--ray band (3--10 keV) and  refer to  \cite{ogle} and \cite{torresi} for a discussion on the spectra at lower energies.\\
A power law is a good parametrization of the hard X--ray continuum (see Table~\ref{tab2}). An inspection of the residuals shows two prominent emission features around 6.4 keV  (Fe K$\alpha$) and 6.9 keV (probably a blend of Fe {\small XXVI} Ly$\alpha$ and Fe K$\beta$ \cite{ogle}), respectively (Figure \ref{spec120}). From the width of the fluorescence iron line ($\sigma$=0.12~keV) we confirm the BLR as the zone where the line is emitted, in agreement with \cite{ogle}.\\

\section{Sample properties}

The X--ray properties of the entire sample, reported in  Table~\ref{tab2}, suggest the following considerations:

\begin{table*}[!]
\begin{center}
\caption{Summary of the X--ray spectral properties of the MAGN sample. The $\gamma$--ray fluxes and photon indices are also reported.}
\begin{tabular}{l | c | c | c | c | c | c | c}
\hline \textbf{Name}  &\textbf{Class}  &\textbf{N$_H$}$^{a}$ &\textbf{F$_x$}$^{b}$ &\textbf{$\Gamma_x$}   &\textbf{Iron line}   &\textbf{F$_\gamma$}$^{c}$  &\textbf{$\Gamma_\gamma$}\\
\\ 
\hline 3C 78/NGC 1218  &FRI    & 0.11$\pm$0.07 & 0.044$\pm$0.01 &2.03$^{+0.27}_{-0.24}$   & no &4.7$\pm$1.8     &1.95$\pm$0.14 \\
 
\hline 3C 84/NGC 1275 & FRI    &-   &0.76$\pm$0.01   &1.78$\pm$0.02    &? &222$\pm$8 & 2.13$\pm$0.02\\

\hline M87/3C 274          & FRI    &0.04$\pm$0.03   &0.03$\pm$0.003    &2.12$\pm$0.13    &no       &24$\pm$6   & 2.21$\pm$0.14\\

\hline NGC 6251 {\it XMM--Newton}   & FRI    & 0.054$\pm$0.014  & 0.36$^{+0.01}_{-0.02}$   &1.89$\pm$0.04  &no  &36$\pm$8  & 2.52$\pm$0.12\\

\hline PKS 0625-354      & FRI    & $<$0.1  & 0.25$\pm$0.01   & 2.53$^{+0.07}_{-0.06}$   &no  &4.8$\pm$1.1  & 2.06$\pm$0.16\\

\hline Cen A                     & FRI    & 8.1$\pm$0.1       & 19$\pm$1  &1.44$\pm$0.03   &yes &214$\pm$12 &2.75$\pm$0.04\\

\hline\hline 3C 120                   & FRI ?   &-  & 4.5$^{+0.5}_{-0.4}$  &1.76$^{+0.03}_{-0.04}$  &yes &29$\pm$17  &2.71$\pm$0.35\\

\hline\hline 3C 111 {\it Suzaku}  & FRII   & 0.79$\pm$0.01  & 2.1$\pm$0.04  &1.58$\pm$0.01   &yes  &40$\pm$8   &2.54$\pm$0.19\\

\hline 3C 111 {\it XMM--Newton} &  -       & 0.78$\pm$0.01  & 4.7$\pm$0.09  &1.63$\pm$0.01   &yes  & -                   &-\\
 
\hline 3C 207                  & FRII/SSRQ    &$<$0.13 & 0.15$^{+0.02}_{-0.01}$ &1.65$^{+0.12}_{-0.08}$ &hint  &24$\pm$4 &2.42$\pm$0.10\\
 
\hline 3C 380                  & FRII/CSS & - &0.15$\pm$0.01   &1.48$\pm$0.09  &hint  &31$\pm$18    &2.51$\pm$0.30\\
\hline
\multicolumn{8}{l}{(a) Intrinsic N$_{\rm H}$ is in units of 10$^{22}$~cm$^{-2}$, in addition to the Galactic column density.}\\
\multicolumn{8}{l}{(b) F$_{\rm x}$ (2--10 keV) is in units of 10$^{-11}$~erg~cm$^{-2}$~s$^{-1}$. Fluxes are unabsorbed.}\\
\multicolumn{8}{l}{(c) F$_{\gamma}$ ($>$100 MeV) is in units of 10$^{-9}$~phot~cm$^{-2}$~s$^{-1}$.}\\
\end{tabular}
\label{tab2}
\end{center}
\end{table*}

\begin{figure}
\includegraphics[width=65mm]{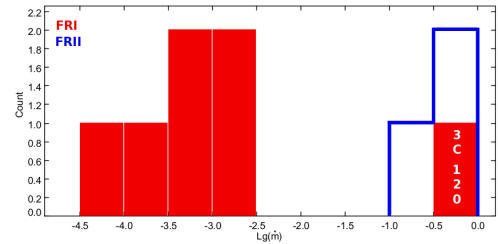}
\caption{Histogram showing the mass accretion rate in Eddington units of the FRI and FRII radio sources belonging to the MAGN sample. The two classes are clearly separated.}
\label{edd}
\end{figure}

\begin{figure}
\includegraphics[width=45mm]{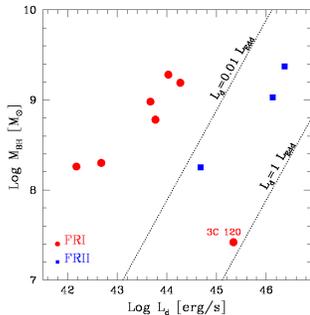}
\caption{Black hole mass vs disk luminosity for the sources of the MAGN sample considered in this work. The dotted lines limit regions with different accretion rates.
The FRI 3C~120 is in the FRII zone, populated by objects with accretion rates ($\dot{m}$=L$_{\rm bol}$/$\epsilon$L$_{\rm Edd}$; $\epsilon$=1) between 0.01 and 1.}
\label{3c120}
\end{figure}

\begin{itemize}
\item FRIIs have, on average, X--ray photon indices harder than FRIs ($\Gamma_{\rm FRI}\sim$2 and $\Gamma_{\rm FRII}\sim$1.6) supporting the idea that different electron populations (non--thermal in FRIs and thermal in FRIIs) are involved in the production of the X--ray continuum (see also \cite{balmaverde}); 

\item the iron line was not detected in four out of six genuine FRIs, supporting the inefficient accretion flow scenario.
In 3C~84, located within the Perseus cluster, there is a claim of an iron line \cite{churazov} whose origin is still uncertain.
In Cen~A the nature of the observed line has been debated.
The low accretion rate of this source ($\dot{m}<$10$^{-3}$) \footnote{On the basis of the SED of quasars we convert L$_{\rm 2-10kev}$ to L$_{\rm bol}$ using a convertor factor $\sim$30 \cite{elvis}. We define $\dot{m}$=L$_{\rm bol}$/$\epsilon$L$_{\rm Edd}$, where the radiative efficiency $\epsilon$ is assumed equal to unity (see \cite{marchesini}).} rejects the hypothesis of an efficient disk, making difficult to explain the presence of a fluorescence line at 6.4 keV.
A possible solution is that the line arises from the interaction between external jet X--ray photons (having the jet a wide opening angle) and surrounding material \cite{paola2003}. Incidentally we note that a structured jet, consisting of an inner fast spine and an external slow layer has been proposed to explain the GeV emission in misaligned FRI AGNs \cite{ghis05};

\item when the accretion disk is efficient and the line well constrained  it is possible to estimate the velocity width and localize the emitting gas. This seems to be located in the BLR, confirming rich in photons circumnuclear environment  in sources with high accretion rates;

\item the X--ray luminosity can be utilized to estimate $\dot{m}$ for each source of the sample.
In Figure~\ref{edd} the histogram of the MAGN accretion rates in Eddington units is shown. 
The bimodal behavior is evident, being FRIs and FRIIs clearly separated.  This is even more clear in Figure~\ref{3c120} where black hole masses  are plotted against disk luminosities (assuming disk luminosity $\simeq$ bolometric luminosity).
We note that in both figures 3C~120 falls well within the region occupied by FRIIs objects, as expected;

\item the environment of FRIs is generally believed to be poor in cold/dusty matter \cite{chiaberge}, \cite{baldi}.
Unfortunately, our limited sample does not allow to support this picture. Indeed, all the MAGN sources, both FRIs and FRIIs, show negligible absorption in excess to the Galactic one. The only exceptions are Cen~A (FRI) and 3C~111 (FRII). Cen~A is known to be bisected by a prominent dust lane that can explain the high column density value. 3C~111 is occulted by a Galactic dark cloud in front of it.
\end{itemize}
 
\section{Conclusions}
The X--ray analysis of MAGNs confirms the presence of different engines in FRI and FRII radio sources, as previously suggested by optical and infrared studies. In particular, FRI sources of the sample have Eddington ratios systematically lower ($\dot{m} <$0.003) than FRIIs ($\dot{m} >$0.1). This is in agreement with the general absence of the iron line in FRIs' X--ray spectra. The peculiar source 3C~120, despite its radio classification, is very similar to 3C~111, the only genuine FRII radio galaxy of the sample (the other two FRIIs, 3C~207 and 3C~380 are quasars).
The iron line clearly detected in 3C~120 and 3C~111 is probably produced in the BLR (rich in photons and matter). The jets of MAGNs having efficient accretion disks seem to propagate in an ambient plenty of photons. This would explain, at least at zeroth order, why External Compton is the dominant process responsible for the high energy photons emission in FRIIs' jet. Instead in FRIs, inefficient accretion and paucity of environment photons make the Synchrotron Self Compton the most important process for the production of $\gamma$--rays.

\bigskip 
\small
\begin{acknowledgments}
ET greatly thanks Paola Grandi for intense and inspiring discussions.
The {\it Fermi}--LAT collaboration is gratefully acknowledged. ET kindly thanks the conference organizers and in particular Roopesh Ojha.
This work was supported by the Italian Space Agency (contract ASI/GLAST I/017/07/0).
\end{acknowledgments}

\bigskip 

\end{document}